\documentstyle[prl,floats,aps,psfig]{revtex}

\begin{document}
\draft

\twocolumn[\hsize\textwidth\columnwidth\hsize\csname @twocolumnfalse\endcsname
\preprint{AEI-2001-009, gr-qc/0102037}

\title{Plunge waveforms from inspiralling binary black holes}

\author{J. Baker, B. Br\"ugmann, M. Campanelli, C. O. Lousto,
and R. Takahashi}

\address{Albert-Einstein-Institut,
Max-Planck-Institut f{\"u}r Gravitationsphysik,
Am M\"uhlenberg 1, D-14476 Golm, Germany}

\date{February 6, 2001}

\maketitle
\begin{abstract}

We study the coalescence of non-spinning binary black holes from near
the innermost stable circular orbit down to the final single rotating black
hole. We use a technique that combines the full numerical approach to solve
Einstein equations, applied in the truly non-linear regime, and linearized
perturbation theory around the final distorted single black hole at later 
times.  We compute the plunge waveforms which present a non negligible 
signal lasting for $t\sim 100M$ showing early non-linear ringing,
and we obtain estimates for the total gravitational energy 
and angular momentum radiated.

\end{abstract}

\pacs{04.25.Dm, 04.25.Nx, 04.30.Db, 04.70.Bw; AEI-2001-009}

\vskip2pc]


The next few years will mark the birth of a new field, gravitational wave 
astronomy.  New extremely sensitive gravitational wave interferometers 
(LIGO and GEO) are nearing completion and should begin taking scientific 
data in about a year to be joined later by VIRGO and the LISA space mission.
The expectation of very strong gravitational wave emissions from the
merger of black hole -- black hole binary systems, and some indirect
indications that these systems may be commonly generated in globular
clusters \cite{Zwart99} makes them one of the most promising
candidates for early observation.

The interpretation of merger events, and in some cases even their
detection, requires a theoretical understanding of the gravitational
waveforms based on general relativity.  Several theoretical approaches
have been developed for treating these systems.  So far the
post-Newtonian (PN) approximation, has provided a good understanding
of the early slow adiabatic inspiral phase of these systems, and
extending PN calculations to the {\it innermost stable circular orbit}
(ISCO) may be possible using resummation techniques
\cite{Buonanno00a}.  In its final moments, though, after the black
holes are closer than ISCO, the orbital dynamics is expected to be
replaced by a rapid {\it plunge} and coalescence. An understanding of
the final burst of radiation coming from the black hole merger can
significantly enhance the detectability of systems with total mass
$M>35M_\odot$ \cite{Damour00a} and even qualitative information is
very important for developing search strategies \cite{Flanagan97b}.
It is generally expected that the dynamics near the ISCO can only be
treated by a fully non-linear simulation of Einstein's equations.

Thus far the numerical treatment of black hole systems has proved
difficult. A key limiting factor is the achievable evolution time
after which the code fails due to numerical problems associated with
the black hole singularities. The first fully $3D$ simulation of
spinning and moving black holes was performed in \cite{Bruegmann97}
for a `grazing collision', where the black holes start out well within
the ISCO. Recently, evolution times of about 9$M$--30$M$ have been
achieved \cite{Brandt00,Alcubierre00b}.
However, these simulations, even
beginning late in the plunge near the onset of quasi-normal ringing
still generate only about two wave cycles.  On the other hand, a third
approach, the `close limit' approximation treating the late-time
ring-down dynamics of the system as a linear perturbation of a single
stationary black hole has shown great effectiveness in estimating the
radiative dynamics once the strong non-linear interactions have
passed, for example in the case of grazing collisions without spin
\cite{Khanna99a}.

In order to provide expectant observers with some estimate of the full
merger waveforms from binary black hole systems `within a factor of
two', and to guide future, more advanced numerical simulations, we
have implemented an interface between full non-linear numerical
simulations and close limit approximations. This interface allows us
to apply the numerical and close limit treatments in sequence, moving
back the finite time interval of full non-linear numerical evolution
to cover the earlier part of the plunge and then computing the
complete black hole ring-down and the propagation of radiation into
the wave zone with a close limit treatment based on the Teukolsky
equation \cite{Teukolsky73}.  In \cite{Baker00b} we have developed the
basic idea for the required interface and applied it to a model case,
the head-on collisions of two black holes producing complete waveforms
for the first time in full $3D$ numerical relativity.

In this letter we present the first theoretical predictions from
non-axisymmetric binary black hole mergers starting from an estimate
of the innermost
stable circular orbit, based on a generalization of \cite{Baker00b} to
include angular momentum. We explicitly derive an astrophysically
plausible estimate for the radiation waveforms and energy which can be
expected from a system of equal mass black holes with no intrinsic
spin. We also estimate the duration of the plunge phase.


Currently no genuinely astrophysical description of ISCO initial data
for numerical simulations exists. As a reasonable starting point we
will use approximate ISCO data based on the effective potential method
of \cite{Cook94} as derived in \cite{Baumgarte00a} for the puncture
construction \cite{Brandt97b} of black hole initial data. The solution
to the general relativistic constraint equations is constructed within
the Bowen-York, conformally flat, longitudinal ansatz.  To locate the
ISCO, the minimum in the binding energy along sequences of constant
(apparent) horizon area is studied as a function of the total angular
momentum of the system.  
For puncture data the ISCO is characterized by the parameters
\begin{eqnarray}
m=0.45M,\ L=4.9M,\ P=0.335M,\ J=0.77M^2,
\end{eqnarray}
where $m$ is the mass of each of the single black holes, $M$ is the
total ADM mass of the binary system, $L$ is the proper distance
between the apparent horizons, $P$ is the magnitude of the linear
momenta (equal but opposite and perpendicular to the line connecting
the holes), and $J$ is the total angular momentum.


Our full numerical evolutions are carried out using the standard
ADM decomposition of the Einstein equations.  The lapse is determined
by the maximal slicing condition, and the shift is set to zero.  Our
biggest runs had $512^2\times256$ grid points and a central
resolution of $M/24$.
We use the Cactus Computational Toolkit to implement these simulations
\cite{Cactusweb}. Our typical evolution times reach $T\approx15M$. After 
this time, the `grid stretching' associated with the singularity
avoiding property of maximal slicing crashes the code. This is
consistent with the evolution time for grazing collisions of about
$30M$, because for `ISCO' runs the time scale for grid stretching is
determined by the individual black holes of mass $0.45M$ that are
present for most of the run, and not by the final black hole of mass
$1M$ that quickly appears in the grazing collisions.


The fundamental problem in determining initial data for the close
limit perturbative approach is to define a background Kerr metric on
the later part of the numerically computed spacetime. We look for a
hypersurface and a spatial coordinate transformation such that
the transformed numerical data on this hypersurface is close to a Kerr
metric in Boyer-Lindquist coordinates to which all perturbative
quantities refer to. The first order gauge and tetrad
invariance of the perturbative formalism implies that the results will
not depend strongly on small variations in our procedure.

We start with the estimate that the background Kerr black hole is
given by the parameters $M$ and $a$ of the initial data. Then we
correct those values with the information about the energy and
angular momentum radiated. This procedure quickly converges.
The slice is
defined by identifying the numerical time coordinate with the
Boyer-Lindquist time, since we have found that in the late stages of
the numerical evolution the maximal slicing lapse approaches quite
closely the Boyer-Lindquist lapse $\left(-g^{tt}_{Kerr}\right)^{-1/2}$
outside the black hole.  To obtain the $r$ coordinate we compute the invariant
${\cal I}=
\tilde C_{\alpha\beta\gamma\delta}\tilde C^{\alpha\beta\gamma\delta}$,
where
$\tilde C_{\alpha\beta\gamma\delta}= C_{\alpha\beta\gamma\delta}+
(i/2)\epsilon_{\alpha\beta\rho\lambda}C^{\rho\lambda}_{\,\,\,\,\gamma\delta}$,
is the self-dual part of the Weyl tensor.  Since
in Boyer-Lindquist coordinates ${\cal I}=3M^2/(r-ia\cos\theta)^6$ for the
Kerr background we can invert this relationship to assign a value of
$r$ to each point of the numerically generated spacetime.  For the
coordinate $\theta$ it turned out to be sufficient to adopt the
corresponding numerical coordinate. Since the binary system carries a
non-negligible amount of angular momentum $J$ producing frame dragging
effects, we supplement the Cartesian definition of $\phi$ with a
correction that makes the $g_{r\phi}$ component of the metric to
vanish (primes denote numerical coordinates)
\begin{equation}
\phi=\phi' + \int (g'_{r'\phi'}/g'_{\phi'\phi'}) dr'.
\end{equation}

The Teukolsky equation is a complex linear wave equation for the two
degrees of freedom of the gravitational field in a Kerr background.
It requires as initial data the Weyl scalar $\psi_4=
-C_{\alpha\beta\gamma\delta}n^\alpha\bar{m}^\beta n^\gamma
\bar{m}^\delta$, with the usual notation for the null tetrad, and its
time derivative $\partial_t\psi_4$ (see \cite{Campanelli98c} for the
ADM decomposition of these quantities).  First, a tetrad is
constructed from the basis vectors of the numerical coordinates, and a
Gram-Schmidt procedure is used to ensure orthonormalization with
respect to the full numerical metric. The tetrad is then rotated to
approximate the required background tetrad. This procedure determines
$\psi_4$ as a linear combination of all the other five Weyl scalars
$\psi_{4,3,2,1,0}$ with coefficients depending on the background
coordinates $r$ and $\theta$, and on $M$ and $a$ \cite{Baker01a}.
With the Cauchy data $\psi_4$ and $\partial_t\psi_4$ extracted in
appropriate coordinates, we can proceed with the evolution by
numerically integrating the Teukolsky equation as described in
\cite{Krivan97a}.  Here one can implement all the desired features for
stable evolution, excision of the event horizon, mesh refinement
through the use of the tortoise coordinate $r^*$, non-vanishing
background shift, imposition of consistent boundary conditions on
$\psi_4$, etc.  The perturbative description is then able to
efficiently follow the evolution of the system forever.

A key question is at what time we can actually make the transition
from full numerical to perturbative evolution. One of the tests we
perform is to compute at every $1M$ of evolution the speciality index
$S = 27 {\cal J}^2/{\cal I}^3$ \cite{Baker00a}, a combination of
curvature invariants $({\cal J}=\tilde C_{\alpha\beta\gamma\delta}
\tilde C^{\gamma\delta}_{\,\,\,\,\rho\lambda}
\tilde C^{\rho\lambda\alpha\beta})$.
For the algebraically special Kerr spacetime $S = 1$, and the size of
deviation from 1 provides a guide on how close a numerical spacetime
is to Kerr. A complementary experiment is to do a full numerical
evolution, during which we extract the data for the Teukolsky equation
every $1M$. Computing the Teukolsky evolution starting at different
times, we can check whether the final results (radiated energies and
waveforms) depend on the time $T$ at which the transition took
place. If the binary system has reached a regime where all further
evolution can be described by the linearized Einstein equations, the
final results should be independent of the choice of the transition
time $T$. This constitutes an important built-in self consistency test 
of our method.

\begin{figure}[t]
\begin{center}
\begin{tabular}{@{}lr@{}}
\psfig{file=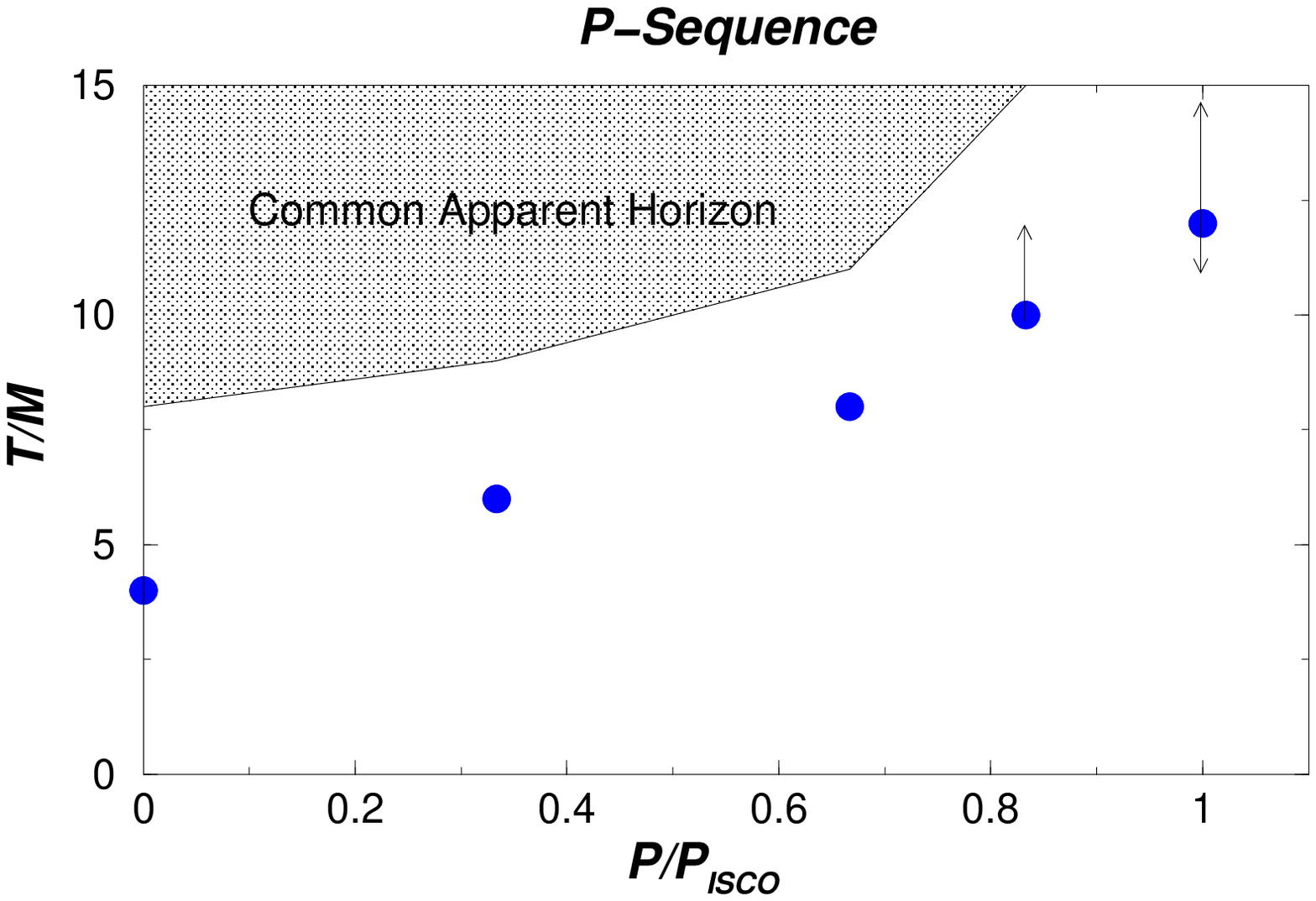,width=2.8in}\\
\psfig{file=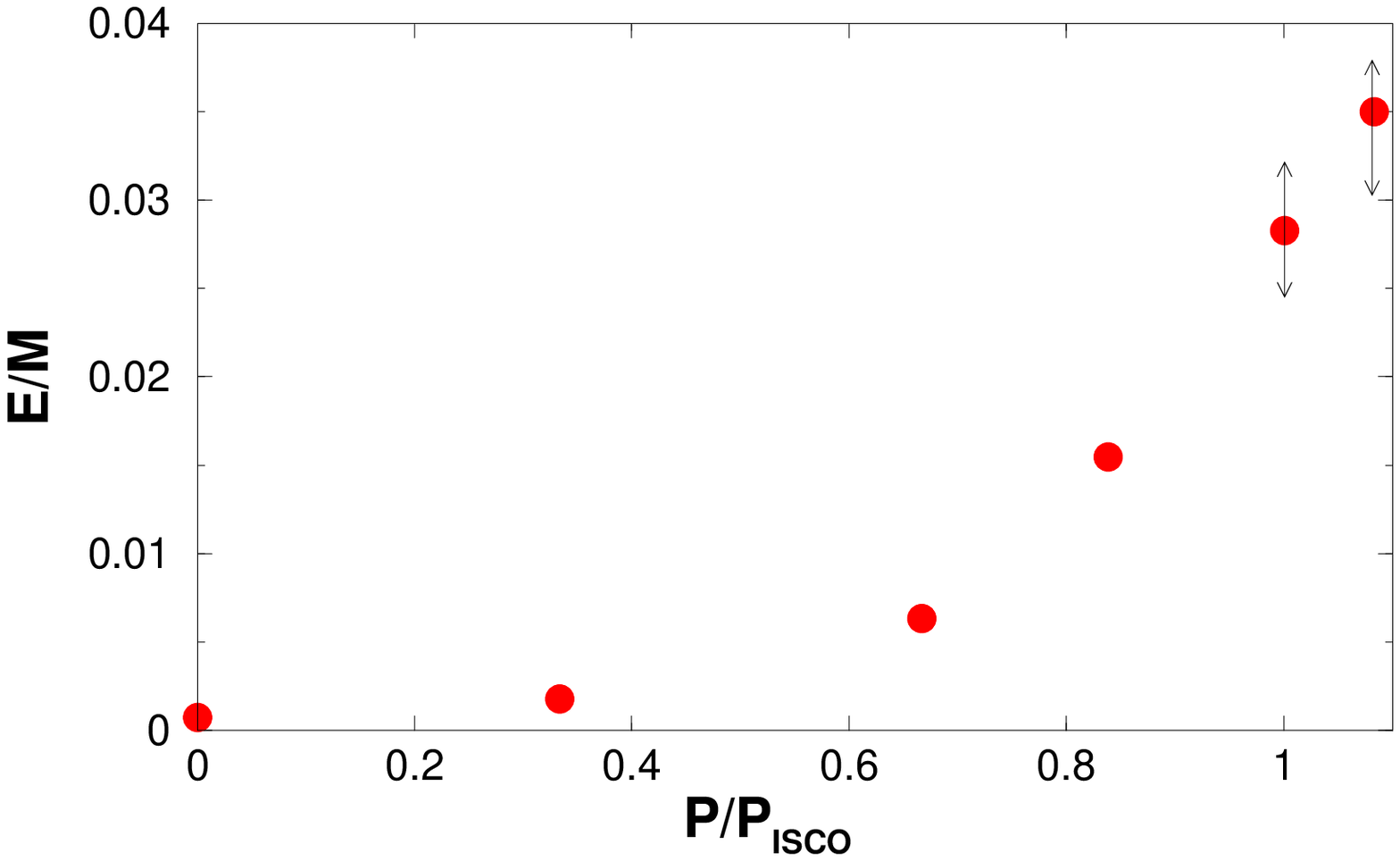,width=2.8in} 
\end{tabular}
\end{center}
\caption{Linearization time and total radiated energy.}
\label{fig:SeqP}
\end{figure}

As mentioned above, we have successfully tested the basic procedure 
for the head-on collision of black holes \cite{Baker00b}.  In
\cite{Baker01a} we report on non-trivial consistency checks, e.g. 
quadratic convergence to vanishing gravitational radiation, which our
method passes for the evolution of Kerr initial data.  Now, in order
to better understand the physics of the plunge we have designed a set
of sequences approaching the ISCO by varying one of its physical
parameters. Several different sequences are possible, which we will
discuss in a longer paper. 
Here we report only on the
`P-sequence' for which we keep the separation constant at
$L=L_{ISCO}=4.9M$, but vary the linear momentum,
$P/P_{ISCO}=0,\frac{1}{3},\frac{2}{3},\frac{5}{6},1,\frac{13}{12}$.
With this sequence we can vary continuously from the head-on
collisions treated effectively in \cite{Baker00b} to the case we are
most interested in, $P = P_{ISCO}$. None of the elements of this
sequence starts in the close limit regime.  Referring to Fig.\
\ref{fig:SeqP}, we observe that the minimal time of full numerical
evolution $T$ needed to switch to perturbation theory initially grows
roughly linearly with increasing momentum, with a further deviation
towards longer times when approaching the ISCO.  We also show the time
at which a common apparent horizon appears, if it actually appears
during the achievable $15M$ of full numerical evolution.  In our
simulations this provides an upper limit to the linearization time.
We note that with the current method $P=P_{ISCO}$ is a marginal
case. Based on the $S$ invariant and the transition time experiment,
linearization for the ISCO happens in the range $T=11M$--$15M$. For
$P/P_{ISCO}=\frac{13}{12}$ the $S$ invariant does not indicate
linearization much before $T = 15M$.  As shown in Fig.\ \ref{fig:SeqP}, the
total radiated gravitational energy grows quadratically with the
momentum $P$ for small values, as one would expect from dimensional
arguments and from extrapolation of close limit results
\cite{Khanna99a}. The error bars are based on variations for different
transition times.

\begin{figure}
\hspace*{3mm}
\psfig{file=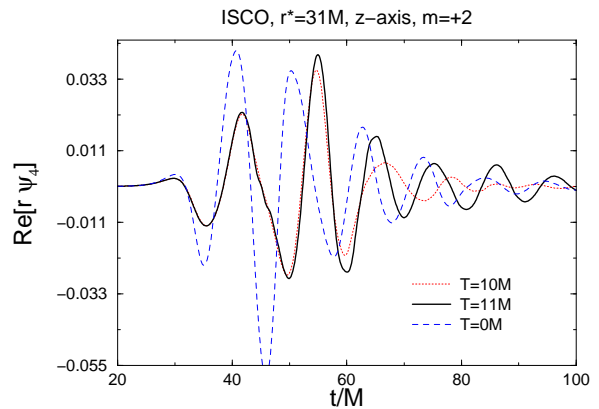,height=2.2in} 
\caption{ISCO waveform for two transition times compared
against the result with no full numerical evolution, $T=0$.}
\label{fig:WF}
\end{figure}

Fig.\ \ref{fig:WF} shows the real part of $\psi_4$ as seen by an
observer located at radial coordinate $r*=31M$ along the orbital pole.
We display the leading $m=2$ mode, since we decompose the Weyl scalar
into $e^{im\varphi}$ modes. The waveforms are obtained for $0$, $10$,
and $11M$ of full numerical evolution plus $120M$ of linear evolution.
The $T = 0$ waveform is displayed only to illustrate the importance of
the full numerical evolution. As is suggested by comparing the
waveforms for $T=10M$ and $11M$, we have precisely determined the
first part of the waveform lasting up to about $t=60M$.  Beyond
that the agreement is less precise.  In the final region
some details of our result, in particular the phase, are still
sensitive to elements of the approximation procedure. Nevertheless, we
judge that we have met our goal of a waveform estimate within a
`factor of two' even in that region. The exact shape of the waveforms
after $t=60M$ may still change when more sophisticated numerical
techniques are applied.

Let us summarize some of the robust features of our waveform.  The
duration of the most significant part of the waveform is roughly
$100M$, lasting for about twice as many wave cycles as in the head-on
collision case.  There is a `ring-up' with increasing amplitude until
about $t\approx60M$ followed by a `ring-down'.  If we define the
duration of the plunge by the time interval from the beginning of the
signal to the onset of the ring-down in Fig.\ \ref{fig:WF}, we obtain
a plunge time of $\approx30M$. This is comparable to the plunge time
derived by different means in PN calculations in the range from $41M$
to $77M$ \cite{Buonanno00a}. For the type of initial data we use the
plunge time is estimated to be equal to the orbital period at the ISCO
in \cite{Cook94,Baumgarte00a}, which is $37M$ and again consistent
with what we find.

For an orbital period of $37M$ we expect to see quadrupole radiation
at half that period.  In fact, a frequency decomposition of the
waveform shows a dominating component with periods close to $12M$
and $19M$, the two most weakly damped quasi-normal
oscillation frequencies for $m=\pm2$ \cite{Leaver85}
of our final black hole with $a=0.8M$. This suggests that at
around $19M$ there is a mixture of quasi-normal ringing and orbital
components.  For a system with total mass $35M_\odot$ our two
principal frequency components correspond to frequencies of roughly
$300Hz$ and $475Hz$, which are within the sensitive range of typical
interferometric gravitational wave detectors.
Note that there is a discrepancy of roughly a factor of two in the
orbital period at the ISCO between the initial data analysis and PN
calculations, i.e. our initial data corresponds to tighter
configurations than are found in PN studies.

Let us point out that the observed time for linearization of less than
$15M$ is significantly shorter than the $\approx30M$ it takes until
ring-down. The onset of linear evolution need not immediately result
in a ring-down waveform. The linearization time is also less than the
time until a common apparent horizon appears, but recall what sets the
range of validity of the close limit approximation. It is not the
presence of a common horizon, but rather that the black holes sit in a
common gravitational well, which occurs earlier. The briefness of the
non-linear phase of the plunge is good news, because this is the
technical reason why we can perform these simulations with the current
techniques.  Still, we want to emphasize that numerical relativity was
essential for achieving these results. For $T=11M$, the
first wave cycle is determined precisely to within $1\%$ of a wave
cycle, while the $T=0M$ waveform with no full numerical evolution 
has a very different appearance and is roughly
90 degrees out of phase.

We estimate the total radiated energy after ISCO to be $\approx3\%$ of
the total ADM mass coming almost entirely from the $m=\pm2$ modes.
This is larger than the $1.4\%$ obtained by extrapolating PN results
\cite{Buonanno00a} and the $1$--$2\%$ obtained by extrapolation of the
close limit \cite{Khanna99a}. The radiated
angular momentum is a delicate quantity to compute involving
correlations of waveforms\cite{Campanelli99}, we estimate the angular
momentum loss to be around $0.2\%$. This confirms the expectation that
not much angular momentum is lost during the plunge and ring-down.


In conclusion, our approach makes it for the the first time possible
to study the fundamentally non-linear processes taking place during
the final plunge phase of the collision of two well-separated black
holes, starting from an estimate for the ISCO location.
The results presented here show that
full 3D numerical evolution is essential to describe the non-linear
interaction of binary black holes. The interface to a linear evolution
of Einstein equations allows us to target the full numerical evolution
where it really matters.

Ours are the first but certainly not the final numerical results for
black hole systems near the ISCO. Further work will be done to extend
the duration of the numerical simulations and to move toward more
astrophysically realistic initial data descriptions, e.g.\ given by an
interface to the post-Newtonian approximation. The technology we have
developed makes it possible to explicitly study the physical
appropriateness of various initial data descriptions, both by
comparison studies, and by moving to more separated `pre-ISCO'
configurations. As another astrophysical application of numerical
relativity one can now begin to characterize the effects of black hole
spins and relative mass differences on the radiation waveforms.

We wish to thank M. Alcubierre, R. Price, B. Schutz, and E. Seidel for
their support and many helpful discussions. M.C. was partially
supported by a Marie-Curie Fellowship (HPMF-CT-1999-00334).  Numerical
computations have been performed at AEI, NCSA, and LZG.

\bibliographystyle{bibtex/prsty}
\bibliography{bibtex/references}
\thebibliography{PRL}

\end{document}